\documentclass{emulateapj}
\slugcomment{Accepted by ApJ 1/16/12}
\usepackage{color}

\shorttitle{Initial Rotation Rates of Massive Stars}
\shortauthors{ROSEN ET AL.}

\begin{document}

\title{What Sets the Initial Rotation Rates of Massive Stars?}

\author{
Anna L. Rosen\altaffilmark{1}, Mark R. Krumholz\altaffilmark{1}, Enrico Ramirez-Ruiz\altaffilmark{1}
}

\altaffiltext{1}{Department of Astronomy and Astrophysics, University of California Santa Cruz, 211 Interdisciplinary Sciences Building, 1156 High Street, Santa Cruz, CA 95064, USA; rosen@ucolick.org}

\begin{abstract}

The physical mechanisms that set the initial rotation rates in massive stars are a crucial unknown in current star formation theory. Observations of young, massive stars provide evidence that they form in a similar fashion to their low-mass counterparts. The magnetic coupling between a star and its accretion disk may be sufficient to spin down low-mass pre-main sequence (PMS) stars to well below breakup at the end stage of their formation when the accretion rate is low. However, we show that these magnetic torques are insufficient to spin down massive PMS stars due to their short formation times and high accretion rates.  We develop a model for the angular momentum evolution of stars over a wide range in mass, considering both magnetic and gravitational torques. We find that magnetic torques are unable to spin down either low or high mass stars during the main accretion phase, and that massive stars cannot be spun down significantly by magnetic torques during the end stage of their formation either. Spin-down occurs only if massive stars' disk lifetimes are substantially longer or their magnetic fields are much stronger than current observations suggest.

\end{abstract}


\section{Introduction}

While there has been significant theoretical attention to understanding the initial rotation rates of Sun-like stars, far less work has been done on more massive stars. Since the stellar evolutionary path depends on the rate of mass loss and internal mixing, both of which are enhanced by rotation \citep{bc93, mm10}, our inability to predict initial rotation rates is a limiting factor in stellar evolution theory. Observations of young, massive stars provide evidence that they form in a similar fashion to their low-mass counterparts: via gravitational collapse of a molecular cloud core \citep{mt03, zph08, dav11}. These cloud cores are slowly rotating but have very large radii, and thus have high initial angular momenta. This has led to the ``angular momentum problem" in which the initial angular momentum of a cloud core is at least three orders of magnitude greater than the resulting star \citep{good93, bod95, lar10} and must be redistributed or removed during collapse. 

Massive stars form in magnetized high-density turbulent gas clumps \citep{cru99} that are characterized by short core collapse times and high time-averaged accretion rates \citep{mt03}. Due to the high angular momentum content of the diffuse gas, material is unable to be directly deposited on to the central object and is instead circularized at a distance far from the star, resulting in a disk \citep{krum07, krum09}. Observations, although rare, confirm that disks form around massive protostars during cloud collapse \citep{cesnat06, cglrev07, chn11} and the accretion onto these disks is regulated at least in part by the magnetic field \citep{vsc10}. Furthermore,  these disks might evolve like those located around young, low-mass stars \citep{chi06}. The disk transfers mass and angular momentum to the central protostar, which acts to spin it up. This transfer of angular momentum, along with contraction of the protostar towards the main sequence, suggests that young stars should be rotating at or near their break up speed, the rotational speed at which the centripetal force at the equator balances gravity. 

\citet{lkk11} found that gravitational torques prohibit a star from rotating above $\sim50\%$ of its break up speed during formation. However, the observed projected rotation rates of young low mass and some massive stars suggest that they rotate at a much lower fraction. Observations of low-mass PMS stars suggest that their rotation periods span a factor of $\sim30$ and approximately half are slow rotators, rotating at about 10\% of their break up speed \citep{har89, hem07}. The observed rotational velocities of massive stars suggest that they are spinning significantly faster than their low-mass counterparts. \citet{ws06} studied a sample of young massive stars  (${\rm M}_\star > 25\; {\rm M}_{\rm \odot}$) and found that their median rotation rate was 20\% of their break up speed. \citet{hgm10} observed the projected rotational velocity distribution of 220 young B stars and found that approximately 53.3\%  are rapid rotators, rotating with a velocity that is at least 40\% of their break up speed. How these initial rotation rates are achieved and their dependence on stellar mass is still an unanswered question.

The physical mechanism responsible for causing young low-mass stars to be slow rotators has received considerable attention over the last three decades. One popular theory is that during the T Tauri phase (experienced by PMS stars with masses less than $\sim \rm{3 \; M_{\odot}}$), when the accretion rate is low, $\dot{ \rm M}_{\rm a} \lesssim 10^{-7} \; {\rm M}_{\rm \odot}\; {\rm yr}^{-1}$ \citep{hdcm06}, the magnetic connection between the star and its accretion disk can transport substantial angular momentum away from the star, resulting in spin rates well below break up in agreement with observations \citep{koe91, ac96}. The fact that T Tauri stars have strong magnetic fields, typically, several hundred G to several kG \citep{jk07}, long contraction timescales after their main assembly, and long accretion disk lifetimes support this spin down scenario \citep{bou07}. However, \citet{mp05a} and \citet{matt10} found that when the stellar magnetic field lines open due to the differential twisting between the star and disk the resulting rotation rates, while still below break up, are higher than those of the slowest rotators.

Magnetic fields have been detected in a small sample  of young and evolved OB stars. These fields are between a few hundred G to several kG and typically have a bipolar topology \citep{don02, wade06, hub08, gru09, mar10}. The origin of these fields is poorly understood, since the envelopes of such stars are radiative rather than convective, excluding the possibility of a Solar-type dynamo effect \citep{moss01}. The favored hypothesis for the presence of magnetic fields in massive stars is that they are fossil fields that were either accumulated or generated during star formation \citep{wfm11}. \citet{awc08} discovered two very young B stars with strong surface magnetic fields. They found that the younger of the two is a rapid rotator and situated in the first half of the PMS phase, whereas the older star, which might already be on the main sequence, is a slow rotator most likely spun down via magnetic torques.

This implies that massive stars likely have strong magnetic fields present during their formation and that these fields, due to coupling with the accretion disk,  may be able to remove a substantial amount of angular momentum from the star, producing spin rates on the zero-age main sequence (ZAMS) well below break up in a similar fashion to their low-mass counterparts. However, massive stars reach the ZAMS very quickly  since they have short thermal equilibrium timescales. They also have higher accretion rates during their formation and their magnetic fields are weaker relative to their stellar binding energy as compared to low mass stars. They likely have shorter disk lifetimes than contracting low mass stars, since their disks are likely to be quickly photo-disintegrated due to their high luminosities \citep{cglrev07}. All of these factors make magnetic spin-down more difficult. In this paper we explore whether the initial spins of massive stars are regulated by the interaction of their accretion disk with the stellar magnetic field. To study this issue we model the angular momentum evolution for both low-mass and massive protostars by considering both magnetic and gravitational torques. We apply the star-disk interaction model developed by  \citet{mp05a} (hereafter MP05) where the stellar magnetic field is connected to a finite region of the accretion disk, and the twisting of the magnetic field lines due to the differential rotation between the star and disk leads to a spin-down torque on the star. 

This paper is organized as follows. In the following section ($\S$2), we give a brief introduction to how the presence of surface magnetic fields during the protostellar phase can extract angular momentum from the star. We describe our stellar angular momentum evolution model, which include a prescription for protostellar evolution and the star-disk interaction, in  $\S$3. We state our results in $\S$4.  Lastly, we discuss our results  in $\S$5.

\section{Magnetic Torques: Theory \& Background}

Protostars embedded in circumstellar disks accrete material from an angular momentum-rich mass reservoir. If the disk is Keplerian the specific angular momentum content of the circulating material, $j=\sqrt{GM_\star r}$, increases outward and the angular velocity increases inwards. The presence of a stellar magnetic field is able to disrupt the disk outside the stellar radius and channel the disk material along field lines. Spin-down torques will be conveyed to the star due to the differential twisting of the field lines threading the accretion disk at radii where the disk rotates at a lower rate than the star. In this section we give simple scaling arguments to demonstrate how spin evolution varies with stellar mass, before proceeding to a more detailed numerical model in $\S$3. The derivation that follows is an oversimplification and ensures maximum spin down via magnetic braking. We include this section for the reader who is unfamiliar with the literature.

The radial extent of the accretion disk can be altered if the protostar has a magnetic field. The magnetic field is able to truncate the disk at the Alfven radius (denoted $R_{\rm A}$) where the magnetic pressure, $B^2/8\pi$, balances the ram pressure, $\rho v^2$, of the infalling material. Assuming the stellar magnetic field is dipolar and the magnetic field axis is aligned with the rotation axis of the star, the z component of the field in the equatorial plane at a distance $r$ from the star is given by 
\begin{eqnarray}
B_{\rm z} = B_{\rm \star} \left(\frac{r}{R_{\rm \star}}\right)^{-3}
\end{eqnarray}
\noindent
where  $B_{\rm{\star}}$ is the magnetic field strength at the stellar surface. The location at which the magnetic pressure is able to truncate the disk, assuming spherical free-fall accretion, is
\begin{eqnarray}
\frac{R_{\rm A}}{R_{\rm \star}}=& 2.26\left(\frac{B_{\rm \star}}{2\, \rm kG}\right)^{4/7}\left(\frac{\dot{M_{\rm a}}}{10^{-7}M_{\rm \odot}\;\rm{yr^{-1}}}\right)^{-2/7} \nonumber \\
&\times \left(\frac{M_{\rm \star}}{M_{ \rm \odot}}\right)^{-1/7}\left(\frac{R_{\rm \star}}{R_{ \rm \odot}}\right)^{5/7}
\end{eqnarray}
\noindent
where $\dot{M}_{\rm a}$ is the accretion rate. In the case of disk accretion, the truncation radius is in general smaller than the value given in equation (2) by a factor of order unity. For simplicity and for the purpose of this section we neglect this factor in the following discussion.

If the stellar magnetic field lines are connected to the disk the differential rotation between the two will cause the field lines to twist in the azimuthal direction inducing torques on the star. The disk co-rotates with the star at the location $R_{\rm co} \equiv \left(G M_{\rm \star}\right)^{1/3} \Omega^{-2/3}_{\rm \star}$ where $\Omega_{\rm \star}$ is the angular velocity of the star. The stellar field lines that connect to the disk outside the $R_{\rm co}$ spin up the disk and spin down the star. If the field lines connect to a significant portion of the disk outside of $R_{\rm co}$ the star can be spun down to a velocity well below its break up speed.

The stellar magnetic field lines threading an annulus of the accretion disk with width $dr$ will exert a torque:
\begin{eqnarray}
d\tau_{\rm m} = B_{\rm \phi} B_{\rm z} r^2 dr
\end{eqnarray}
\noindent
where $B_{\rm \phi}$ is the azimuthal component of the field generated by the twisting of the field lines relative to the star and is given by
\begin{eqnarray}
B_{\rm \phi} = B_z \frac{\Omega(r) -\Omega_\star}{\Omega(r)}
\end{eqnarray}
where $\Omega$ is the angular velocity of the Keplerian accretion disk. Integrating equation (3) from $R_{\rm A}$ to infinity the total torque on the star due to the stellar magnetic field lines connected to the disk is

\begin{eqnarray}
\tau_{\rm m} = \frac{B^2_{\rm \star} R_{\rm \star}^6}{3} \left(R^{-3}_{\rm A} - 2 R^{-3/2}_{\rm co} R^{-3/2}_{\rm A} \right).
\end{eqnarray}

\noindent
The accretion of disk material at $R_{\rm A}$ adds angular momentum to the star at a rate
\begin{eqnarray}
\tau_{\rm a} = \dot{M}_{\rm a} \sqrt{G M_{\rm \star} R_{\rm A}}. 
\end{eqnarray}
Notice that equation (5) contains both spin-up and spin-down torques acting on the star due to field lines connected to the disk within and outside of $R_{\rm co}$, respectively. In order for the net magnetic torque to transport angular momentum  away from the star (i.e., $\tau_{\rm m}<0$) $R_{\rm A}$ must be greater than
\begin{eqnarray}
R_{\rm A,\,min} \approx 0.63 R_{\rm co}. 
\end{eqnarray}

In a system where the stellar parameters ($M_\star$, $R_\star$, $B_\star$, $\dot{M}_{\rm a}$) are relatively constant there exists an equilibrium state, called the ``disk-locked" state \citep{koe91, ac96, mp05a}, in which the stellar spin rate will adjust to its equilibrium value (i.e., when $\tau_a + \tau_m = 0$). Setting $\tau_{\rm a} = -\tau_{\rm m}$  the equilibrium spin rate, as a fraction of the break-up speed ($ \Omega_{\rm bu}=  \sqrt{G M_{\rm \star} / R_{\rm \star}^3} $), is
\begin{eqnarray}
\frac{\Omega_{\rm \star,eq}}{\Omega_{\rm bu}}=&\frac{1}{2}\left(\frac{R_{\rm A}}{R_{\rm \star}}\right)^{-3/2}\left[0.014\left(\frac{M_{\rm \star}}{M_{\rm \odot}}\right)^{1/2} \left(\frac{\dot{M}_{\rm a}}{10^{-7}M_{\rm \odot}\,{\rm yr}^{-1}}\right) \right. \nonumber \\ 
& \left. \times \left(\frac{B_{\rm \star}}{2\, \rm{kG} }\right)^{-2} \left(\frac{R_{\rm A}}{R_{\rm \star}}\right)^{7/2}+1\right]. 
\end{eqnarray}
\noindent
Assuming that the moment of inertia of the star stays constant, the characteristic timescale to reach equilibrium is: 
\begin{eqnarray}
t_{\rm \star,eq} = k^2 M_{\rm \star} R^2_{\rm \star} \left(\frac{\Omega_{\rm \star,eq} -\Omega_{\rm \star}} {\tau_{\rm a} +\tau_{\rm m}} \right)
\end{eqnarray}
\noindent
where $k$ is the dimensionless radius of gyration whose value depends on the stellar structure. Equation (7) only holds when $R_{\rm A} > R_{\rm \star}$, which is true if the star has a surface magnetic  field strength above a minimum value:
\begin{eqnarray}
B_{\rm \star} > & 400 \left( \frac{ \dot{M_a}}{10^{-7} \; \rm M_\odot \; yr^{-1}} \right)^{1/2} \nonumber \\
	 &\times \left(\frac{M_\star}{M_\odot}  \right)^{1/4} \left(\frac{R_\star}{R_\odot}  \right)^{-5/4}\;G.
\end{eqnarray}
\noindent

\begin{figure}[!t]
\includegraphics[trim=0cm 0cm 1cm 0cm, clip=true, width=0.48\textwidth]{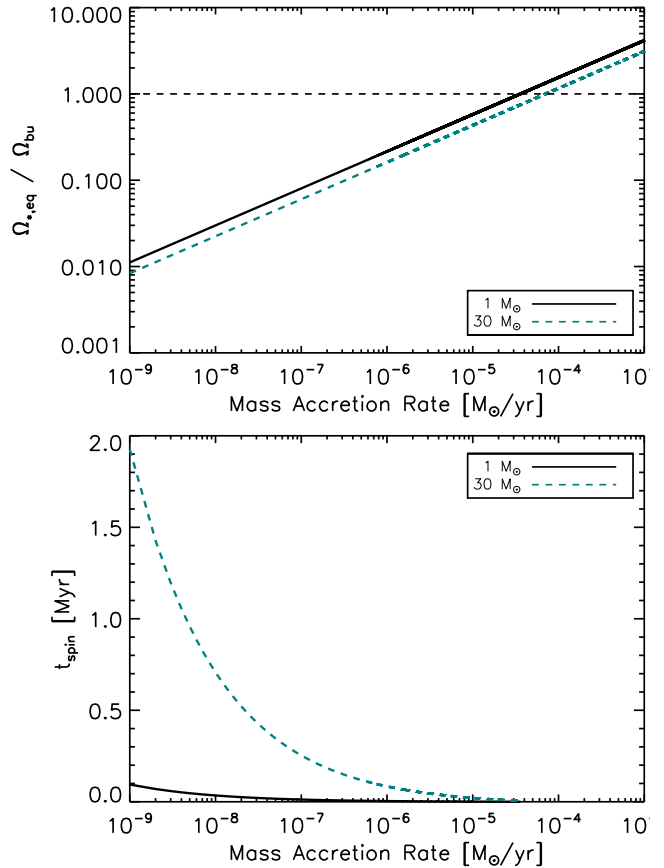}
\centering
\caption{The equilibrium spin rate of a star as a fraction of its break up spin rate (top panel), and the corresponding spin down time scale (bottom panel) for 1 $M_\odot$ (black solid line) and 30 $M_\odot$ (teal dashed line) stars to reach equilibrium. Both stars have a surface magnetic field strength of 2 kG with a dipolar topology and are initially rotating at break up. The horizontal line in the top panel shows where the equilibrium spin rate is equal to the break up rate.}
\label{fig:teq}
\end{figure}

Figure~\ref{fig:teq} shows the equilibrium spin rate as a fraction of the star's break up speed and the corresponding time scales required for a 1 $\rm{M_\odot}$ star and a 30 $\rm{M_\odot}$ star to reach equilibrium starting from rotation at break up, both as a function of the accretion rate. We adopt surface magnetic field strengths of 2 kG similar to observations \citep{wade06, jk07, gru09} and assume $k=0.27$ for a radiative star (e.g., n=3 polytrope). We adopt radii of 3 $\rm R_\odot$ for the 1 $\rm M_\odot$ star (the typical radius of a T Tauri star of this mass) and 7.76 $\rm R_\odot$ for the 30 $\rm M_\odot$ star (ZAMS value). We consider only accretion rates where the equilibrium spin rate is below the break up rate. As the accretion rate increases, the equilibrium spin rate approaches the  break up rate and the equilibrium timescale quickly decreases. We find that magnetic torques produce equilibrium spin rates below break up only for accretion rates below $\rm{\dot{M}_a} \lesssim 5 \times 10^{-5}\; \rm{M_\odot \; yr^{-1}}$, regardless of the stellar mass. In this regard, low- and high-mass stars are similar. The typical mass accretion rates during the main accretion phase, where the majority of the stellar mass is accreted, for low- and high-mass star formation are $5\times 10^{-6} \;  \rm{M_\odot \; yr^{-1}}$ \citep{shu77} and  $5\times 10^{-4} \; \rm{M_\odot \; yr^{-1}}$ \citep{mt03}, respectively. For our adopted field strength, $R_{\rm A}$ for the 30 $\rm M_{\odot}$ star is within the stellar surface at this accretion rate. In contrast, the disk is truncated very close to the stellar surface for the 1 $\rm M_\odot$ star, leading to an equilibrium spin rate close to break up. We conclude that disk truncation does not occur for massive stars and is unimportant for low-mass stars during the main accretion phase. At the lower accretion rates that are likely to occur after the main accretion phase ends, we find that low- and high-mass stars differ in that the latter have much longer equilibration timescales than the former due to their larger inertia. For example, the equilibration timescale for the 30 $\rm{M_\odot}$ star for very low accretion rates is a significant fraction of its stellar lifetime, $\rm{t_{ms}}$ =  5.9 Myr \citep{phm03}. Furthermore, at high accretion rates this timescale is comparable to the star's formation timescale \citep{mt03} suggesting that massive stars are unable to reach spin equilibrium. To further explore the consequences of this analysis we follow the angular momentum evolution of massive protostars to determine the physical conditions that are required to spin them down by magnetic torques.

\section{Stellar Angular Momentum Evolution Model}
The goal of this work is to determine if the initial rotation rates of massive stars can be regulated by magnetic torques due to the interaction of the stellar magnetic field and surrounding accretion disk during formation. To this end, we construct a simple model to track the mass, radius, and angular momentum content of accreting protostars subjected to gravitational and magnetic torques. We describe the elements of this model in the following subsections.

\subsection{Protostellar Model}
We monitor the spin and angular momentum evolution by following the protostellar radius and internal structure evolution during its formation with the use of the one-zone model of \citet{mt03} (hereafter MT03) as updated by \citet{off09}. By treating the protostar as an accreting polytrope and requiring conservation of energy, the evolution of the protostellar radius is given by:
\begin{eqnarray}
\frac{dR_{\rm \star}}{dt} =& \frac{2\dot{M}_{\rm a} R_{\rm \star}}{M_{\rm \star}} \left( 1- \frac{1-f_{\rm k}}{a_{\rm g} \beta_{\rm P}} + \frac{1}{2} \frac{d\log \beta_{\rm P}}{d \log M_{\rm \star}} \right) \nonumber \\
&- 2 \left( \frac{R^2_{\rm \star}}{G M^2_{\rm \star}}\right) \left(L_{\rm int} + L_{\rm I} -L_{\rm D} \right)
\end{eqnarray}
where $\dot{M}_{\rm a}$ is the accretion rate onto the protostar, $f_{\rm k}$ is the fraction of kinetic energy of the infalling material that is radiated away, $\beta_{\rm P}$ is the ratio of radiation pressure to the total pressure, $a_{\rm g} = 3/\left(5-n\right)$ is the coefficient describing the binding energy of a polytrope, $L_{\rm int}$ is the internal stellar luminosity, $L_{\rm I}$ is the rate of energy required to dissociate and ionize the infalling material, and $L_{\rm D}$ is the rate at which energy is supplied from burning deuterium \citep{nhmy00}. The model also includes a few discontinuous changes in polytropic index and radius to represent events such as the onset and cessation of core deuterium burning and the formation of a radiative core. We use the model parameters recommended by \citet{off09} which are based on the detailed stellar evolution calculations by \citet{ho09}.  We refer the reader to MT03 and Appendix B of \citet{off09} for a detailed description of the model and protostellar evolutionary states. 

We treat the protostar as a solid body to follow its angular momentum content ( $J_{\rm \star} = I_{\rm \star} \Omega_{\rm \star}$). We evolve the stellar angular momentum content by computing  the net torque on the star due to the coupling of the stellar magnetic field with the surrounding accretion disk described in $\S$3.3.

\subsection{Accretion History}
The accretion history of our protostars is divided into two distinct accretion phases. The first is the main accretion phase given by the turbulent core model from MT03, which describes an accelerating accretion rate, where the majority of the stellar mass is accreted. This model assumes that the star-forming core is marginally unstable, massive, and supported by turbulent motions. Next, we follow the disk clearing phase in which the accretion disk is no longer being fed by the core envelope. These accretion phases are described in $\S$3.2.1 and $\S$3.2.2.

\subsubsection{Primary Accretion Phase: Core Collapse}
We model the mass accretion using the two-component core model of MT03 which assumes the central region of a molecular cloud core is dominated by thermal motions and the core envelope is dominated by non-thermal motions \citep{mf92}. This leads to a density distribution that is equivalent to the sum of a singular polytropic sphere and a singular isothermal sphere:

\begin{eqnarray}
\rho = \rho_{\rm s} \Biggl(\frac{R_{\rm core}}{r}\Biggr)^{k_{\rm \rho}} + \frac{c^2_{\rm th}}{2\pi G r^2}
\end{eqnarray}

\noindent
where $\rho_{\rm s}$ is the density at the surface of the core,  $R_{\rm core}$ is the core radius, and $c_{\rm th}$ is the thermal sound speed within the core and is assumed to be constant. We adopt the fiducial value of $k_{\rm \rho} = 1.5$ from MT03 in agreement with observations describing the turbulence-supported density profile of massive star forming cores \citep{cm95, vdt00, bsmm02}. 

The accretion rate onto the disk, which is supplied by the background core, is:

\begin{eqnarray}
\dot{M}_{\rm a} \simeq & \frac{\phi_{\rm \star} M_{\rm \star, f} }{t_{\rm \star, ff}} \Biggl[ \Biggl( \frac{M_{\rm \star}}{M_{\rm \star, f}} \Biggr)^{\rm 2j}
 +  \Biggl( \frac{\phi_{\rm \star, th}}{\phi_{\rm \star, nth}} \Biggr)^{2} \nonumber \\
 &\times\Biggl( \frac{\epsilon_{\rm{core}} M_{\rm th} }{M_{\rm \star, f}} \Biggr)^{\rm 2j}  \Biggr]
\end{eqnarray}

\noindent
where $t_{\rm \star, ff} = \bigl(3\pi/32 G \rho \bigr)^{1/2}$ is the free fall time evaluated at $R_{\rm{core}}$, $M_{\rm \star}$ is the current stellar mass, $M_{\rm \star, f}$ is the final stellar mass, and 
\begin{eqnarray}
j=\frac{3\bigl(2 - k_{\rm \rho} \bigr)}{2\bigl(3 - k_{\rm \rho} \bigr)}. 
\end{eqnarray} 

\noindent
The dimensionless constants $\phi$, $\phi_{\rm \star, th}$, and $\phi_{\rm \star, nth}$ are of order unity and depend on $k_{\rm \rho}$ and the magnetic field strength. The efficiency factor, $\epsilon_{\rm core}$, describes how much of the core mass will end up in the star rather than being ejected by the protostellar outflow and we adopt the value of 0.5 from MT03, which is typical of both low-mass \citep{mm00} and high-mass star formation \citep{ckk11}. The parameter $M_{\rm th}$ describes the mass below which the thermal density distribution dominates. For a core with surface density $\Sigma=M_{\rm \star,f} \epsilon_{\rm core}^{-1}/{\pi R^2_{\rm core}}$,  $M_{\rm th}$ is defined as:

\begin{eqnarray}
M_{\rm th} =& 1.23 \times 10^{-3} \left(\frac{T}{20\;K}\right)^3 \nonumber \\ 
& \times \left(\frac{30\epsilon_{\rm core} M_{\rm \odot}}{M_{\star, f} }\right)^{1/2} \Sigma_{\rm 0}^{3/2} M_{\rm \odot}
\end{eqnarray}
\noindent
where $\Sigma_{0} = \Sigma / \rm{\left( 1\;g\;cm^{-2}\right)}$. We further assume that the accretion rate onto the disk is the same as that onto the star and use this value for our protostellar accretion rate. 

\begin{figure}[!t]
\includegraphics[angle = 90,trim=0cm 0cm 1cm 0cm, clip=true, width=0.51\textwidth]{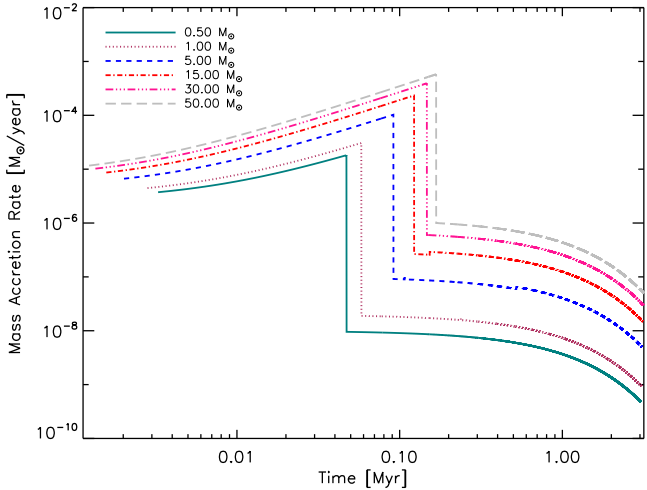}
\centering
\caption{Accretion history of protostars with final masses of 0.5 - 50  $\rm M_{\rm \odot}$, following equations (13) and (16), for our fiducial parameters given in table 1 in $\S$4.}
\label{fig:mdot}
\end{figure}

\subsubsection{Secondary Accretion Phase: Disk Clearing}
 
Late in the formation the core envelope will exhaust its reservoir of mass and no longer feed the accretion disk. We assume that we are left with a thin, Keplerian accretion disk that continues to transfer mass and angular momentum to the central protostar. For simplicity and because observations of disks located around massive stars are very limited, we assume that this results in a decreasing accretion rate as a function of time which we model as a decaying exponential \citep{cc93, yi94, yi95, matt10}: 
\begin{eqnarray}
\dot{M}_{\rm a} = \frac{M_{\rm D}}{t_{\rm a}} e^{-t/t_{\rm a}}
\end{eqnarray}

\noindent
where $M_{\rm D}$ is the remaining mass in the accretion disk (i.e., the total amount of mass that would accrete from $t=0\rightarrow \infty$) and $t_{\rm a}$ is the decay timescale. Since $M_{\rm D}$ and  $t_{\rm a}$ are highly unconstrained, we experiment with different values in $\S$4. Figure~\ref{fig:mdot} shows the accretion history, including both the core collapse and disk clearing accretion phases, for stars with final masses of 0.5-50 $\rm{M_{\odot}}$.

\smallskip

\subsection{Star-Disk Interaction Model}
In $\S$2 we showed how the presence of a stellar magnetic field can remove angular momentum from the star as it accretes matter from an accretion disk. This description assumed that the stellar field lines were connected at all radii of the disk larger than $R_{\rm A}$.  However, the differential rotation between the star and disk will twist the connected field lines. This twisting will cause the magnetic field to undergo a rapid inflation leading to an opening of the field lines, effectively decreasing the size of the disk region that is connected to the stellar magnetic field \citep{love95,ukl02, mp05a}. We now include this effect when calculating the net magnetic  torque on the star with the use of the model developed by MP05, which is an extension to the disk-locking model first developed by \citet{gl78} for accreting neutron stars and extended by \citet{koe91} to describe the star-disk coupling for magnetized T Tauri stars. 

\subsubsection{Magnetic Coupling to the Disk and the Connection State}

The effect of the opening of the magnetic field lines depends on the strength of the magnetic coupling to the disk and how strongly the field lines can be twisted until they are severed. The variable $\gamma(r) = B_{\rm \phi} / B_{\rm z}$ describes the twisting of the magnetic field between the star and disk. This twisting occurs rapidly so a steady state configuration depends on how well the field couples to the disk (i.e., the balance between the differential rotation and the tendency for the magnetic field to untwist).  \citet{ukl02} describe this coupling by a dimensionless magnetic diffusivity parameter, 

\begin{eqnarray}
\beta \equiv \frac{\eta_{\rm t}}{H v_{\rm k}}
\end{eqnarray}

\noindent
where $\eta_t$ is the effective magnetic diffusivity and is of the order of magnitude of the disk's effective viscosity \citep{love95}, $H$ is the scale height of the disk, and $v_k$ is the Keplerian rotation velocity. MP05 assume $\beta$ is constant throughout the disk. The field is strongly coupled to the disk for values of $\beta < 1$ and weakly coupled for $\beta > 1$. \citet{ukl02} find that when $\gamma$ exceeds a value of order unity (defined by the critical twist parameter  $\gamma_c$) the magnetic field will be severed because the magnetic pressure force associated with $B_\phi$ will push outward and cause the dipole field loops to open. The magnetic field is connected to the disk only in the location where $|\gamma| \le \gamma_c$. MP05 use the values $\beta=0.01$ and $\gamma_c = 1$ in their models, and we adopt the same fiducial values in this work. They suggest that $\beta=0.01$ is the most probable value for a T Tauri accretion disk with the use of an $\alpha$ model prescription; however it is uncertain that disks surrounding massive stars will have this same value. For example, massive stars emit more ionizing radiation which will yield a higher ionization fraction on the disk surface, causing $\beta$ to decrease, but these disks are also more massive than those surrounding low-mass PMS stars and are therefore thicker, causing $\beta$ to increase. To account for our uncertainty in this parameter we  experiment with different values in the following section.

MP05 show that the magnetic connection between the star and disk changes at a threshold value of the stellar spin rate. Specifically, the stellar magnetic field will only be connected to a small region of the disk within $R_{\rm co}$ if the stellar rotation rate as a fraction of break up,
\begin{eqnarray}
f = \frac{\Omega_{\rm \star}}{\Omega_{\rm bu}} = \Omega_{\rm \star} \sqrt{\frac{R^3_{\rm \star}}{G M_{\rm \star}}}, 
\end{eqnarray}
\noindent
 falls below:

\begin{eqnarray}
f < \bigl( 1 - \beta \gamma_c \bigr) \bigl( \gamma_{\rm c} \psi \bigr),
\end{eqnarray}

\noindent
where
\begin{eqnarray}
\psi \equiv \frac{2 B^2_{\rm \star} R_{\rm \star}^{5/2} } {\dot{M}_{\rm a} \sqrt{G M_{\rm \star}}}
\end{eqnarray}

\noindent
is a dimensionless parameter that relates the strength of the magnetic field to the accretion rate. This connection state, which MP05 denote as state 1, will result in no spin-down torques transferred to the star. If $f$ exceeds this value then the system is in state 2 which is characterized by a magnetic connection on either side of $R_{\rm co}$ resulting in both spin-up and spin-down torques acting on the star.

\begin{figure}[!t]
\includegraphics[angle = 90,trim=0cm 0cm 0.25cm 0cm, clip=true, width=0.52\textwidth]{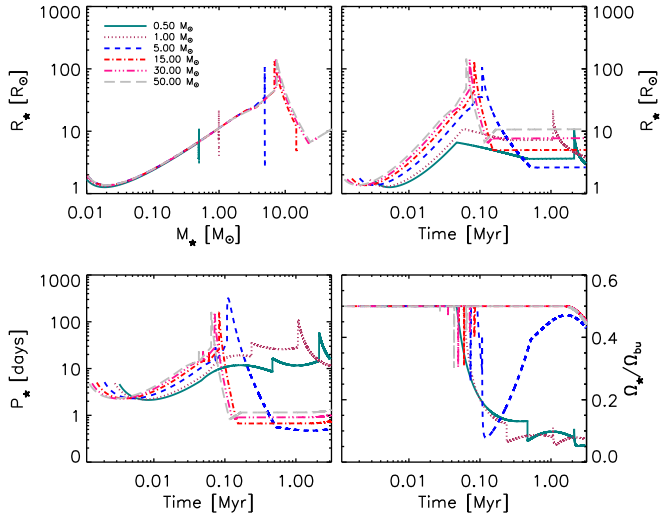}
\centering
\caption{The top left panel shows the stellar radius as a function of stellar mass for stars with masses 0.5 - 50  $\rm M_{\rm \odot}$. The other panels show the stellar radius (top right), stellar period (bottom left), and stellar spin rate as a fraction of break up (bottom right)  as a function of time for stars with masses 0.5 - 50  $\rm M_{\rm \odot}$. Figure~\ref{fig:mdot}  shows the accretion histories.}
\label{fig:stars}
\end{figure}

\subsubsection{Magnetic and Accretion Torques}
The twisting of the magnetic field by the differential rotation between the star and disk causes torques to be conveyed between the two. The twisting of the magnetic field within $R_{\rm co}$ leads to spin-up torques whereas the field lines connected to the disk outside of $R_{\rm co}$ act to spin down the star. If the magnetic field is strong enough then the disk will be disrupted by the stellar magnetosphere where the magnetic stress is able to maintain the accretion rate within the disk. At this location, denoted by $R_{\rm t}$,  the magnetic stress is large enough to remove the excess angular momentum and funnel the disk material along the magnetic field lines. This material and its angular momentum is transferred to the star. If $R_{\rm t} > R_{\rm co}$ the magnetic stress hinders the accretion rate.

The location of $R_{\rm t}$ depends on the connection state of the system. In state 1 the truncation radius is
\begin{eqnarray}
R_{\rm t} = \left(\gamma_{\rm c} \psi \right)^{2/7} R_{\rm \star}.
\end{eqnarray}
\noindent
In state 2 the truncation radius is given by

\begin{eqnarray}
\left(\frac{R_{\rm t}}{R_{\rm co}} \right)^{-7/2} \left[1 - \left(\frac{R_{\rm t}}{R_{\rm co}} \right)^{3/2} \right] = \frac{\beta}{\psi f^{7/3}}. 
\end{eqnarray}
We assume the accreted disk material is quickly integrated into the structure of the star and adds angular momentum to the star at a rate given by equation (6) where $R_{\rm A}$ is replaced by $R_{\rm t}$. This material acts to spin up the star.  

The magnetic connection over a range in radii in the disk can extract angular momentum from the star and transfer it to the disk. If the system is in state 2 then the magnetic field is connected to the disk from $R_{\rm t}$ to $R_{\rm out} = \left(1+\beta \gamma_{\rm c} \right)^{2/3} R_{\rm co}$ which yields a net magnetic torque on the star:

\begin{eqnarray}
\tau_{\rm m}  =& \frac{B^2_{\rm \star} R^6_{\rm \star}}{3\beta R^3_{\rm co}} \left[ 2\left(1+\beta \gamma_{\rm c} \right)^{-1} -  \left(1+\beta \gamma_{\rm c} \right)^{-2}  \right. \nonumber \\ & \left. - 2 \left(R_{\rm co} / R_{\rm t}\right)^{3/2} + \left(R_{\rm co} / R_{\rm t} \right)^{3}  \right].
\end{eqnarray}

\noindent
If the system is in state 1 then the magnetic field is connected to only a small portion of the disk which leads to a negligible torque on the star, so we set $\tau_{\rm m} = 0$ following \citet{matt10}. Note that equation (23) reduces to equation (5) for the limiting case of no field opening ($\gamma_{\rm c} \rightarrow \infty$), marginal coupling ($\beta=1$),  and a disk that is truncated at the Alfven radius ($R_{\rm A}$) and extends to infinity.

\begin{table}[ht!]
\begin{center}
	\begin{tabular}{ | c | c | }
	\hline	
	Parameter             & Fiducial Value \\
	\hline
	$\Sigma$	&   1 $\rm{g\; cm^{-2}}$\\
	$M_{\rm D}$ &	0.02 $M_{\rm \star,f}$ \\
	$t_{\rm a}$ & $10^6$ yr \\
	$B_{\rm \star}$ & 2 kG \\
 	$\beta$ & 0.01 \\
	 $\gamma_{\rm c}$ & 1 \\
 	\hline
	\end{tabular}
\end{center}
\caption{Table of fiducial values used for our model parameters.}
\label{tab:param}
\end{table}

\section{Results}
The initial star-forming core properties are determined by the core mass ($M_{\rm core}$), core density profile ($k_{\rm \rho}$), and core surface density ($\Sigma$). These parameters control the accretion rate for the primary accretion phase as described in $\S$3.2.1.  We initially create a ``pre-collapse" object with a mass less than ${\rm 0.01\;M_\odot}$ which grows in mass with the accretion rate given by equation (13). When the object reaches a mass of 0.01 $\rm{M_\odot}$ we initialize our protostellar and angular momentum evolution model and assume the protostar is initially rotating at 1\% of its break up speed. When the protostar is initialized, it is immediately spun up since the accretion rate is large, so our chosen value for the initial rotation speed is unimportant. We solve equation (11) with the fourth-order Runge-Kutta scheme of \citet{nr07} and update the angular momentum of the star by computing the net torque on the star arising from the accretion and magnetic torques described in $\S$3.3.2. We use this result to update $\Omega_\star$. We cap the stellar rotation rate at 50\% of breakup, a limit imposed by gravitational torques \citep{lkk11}. The fiducial values used for our model parameters are given in table ~\ref{tab:param}.

\subsection{Effect of the Star-Disk Magnetic Interaction} 
Figure~\ref{fig:stars} shows the radial and rotational evolution for stars ranging in final stellar mass from 0.5 - 50  $\rm M_{\rm \odot}$. These models were simulated with the fiducial parameters given in table  ~\ref{tab:param}. The disk-clearing accretion phase is assumed to last 3 Myr, although as discussed in $\S$ 1, this assumption is almost certainly not correct for high mass stars. As we show below, using a shorter disk clearing timescale for the massive stars would only strengthen our results. We choose to run the disk clearing phase for three decay time scales because accretion disks around low-mass stars survive for several million years \citep{hem07}, with an accretion rate that likely decreases with time. The swelling in radius by a factor of three, shown in the upper plots of Figure~\ref{fig:stars}, is a result of the star transitioning from a convective to radiative core \citep{ho09}, which redistributes entropy within the star. For the stars presented in Figure~\ref{fig:stars} this occurs in the primary accretion phase for the most massive stars ($M_{\rm \star,f} \ge 15\;\rm{M_\odot}$) and during the disk clearing accretion phase for the 0.5, 1, and 5 $\rm M_\odot$ stars. If the jump in radius occurs during the main accretion phase, it causes the star to immediately slow down, but the star is almost instantly spun back up because of the high accretion rate. In the case of the 5 $\rm M_\odot$ star, this jump in radius also significantly decreases the spin rate of the star, but since it occurs when the accretion rate is much lower the star only gradually spins up as it contracts and accretes material. In contrast, for the 0.5 and 1 $\rm M_\odot$ stars magnetic torques are able to continue to spin down the star after the jump in radius occurs. We note that \citet{matt10} produced Sun-like stars with faster rotation rates ($\sim20-40\%$ of break up) performing a similar analysis. We report a lower rotation rate for our 1 $\rm M_\odot$ protostar because it has a different radial history than the stars produced by \citet{matt10}. Our 1 $\rm M_\odot$ protostar contracts more slowly than the 1 $\rm M_\odot$ protostar model used by \citet{matt10}. After 3 Myr, our model gives a radius of $3.8$ $\rm R_\odot$ as compared to \citet{matt10}'s  $\sim3$ $\rm R_\odot$. At times \textless 1 Myr, the model radii can differ by factors of $\sim$2. The larger radii in our model produce more spin-down. The differences in predicted radii likely arise because our model accounts for the extra entropy provided both by deuterium burning and by ongoing accretion, while \citet{matt10}'s does not. We do warn, however, that there are significant uncertainties in how much of the accretion entropy is actually absorbed by the star, and differing assumptions on this point can produce significant differences in radial evolution \citep{hok11}.

We find that the torques that arise from the star-disk magnetic interaction are unable to spin down both low-mass and massive protostars during the main accretion phase, but are important during the disk clearing phase, especially for low-mass stars. Low-mass stars begin to spin down the instant the disk clearing accretion phase begins whereas it takes approximately 2 Myr to begin to spin down massive stars for our chosen fiducial values. This suggests that  massive stars are difficult to spin down due to their larger inertia and because their magnetic fields are weaker relative to their stellar binding energy as compared to low mass stars.

\begin{figure}[!t]
\includegraphics[angle = 90, trim=0cm 0cm 2cm 0cm, clip=true, width=0.52\textwidth]{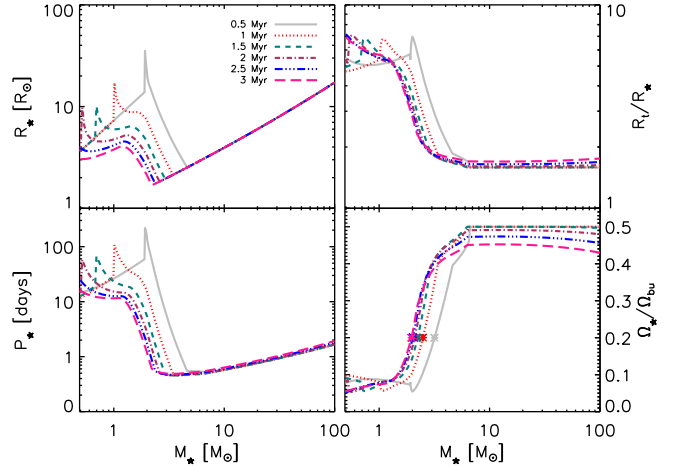}
\centering
\caption{Snapshots of the stellar radius (upper left), disk truncation radius (upper right), stellar period (lower left), and rotation rate as a fraction of break up (lower right) as a function of stellar mass taken at different times during the disk clearing phase for our fiducial case. The times in the legend represent the time that has elapsed since the disk clearing phase began. The points located in the bottom right panel represents the minimum mass of stars rotating at $\gtrsim20$\% of its break up speed. We use this as an indicator of the transition between slow and fast rotators.}
\label{fig:varytdisk}
\end{figure}

Figure~\ref{fig:varytdisk} shows snapshots of the stellar radius, disk truncation radius, stellar period, and stellar rotation rate as a fraction of break up as a function of stellar mass taken at different times during the disk clearing phase. First consider the upper left panel, showing radius versus mass at different times. The $R-M$ relation toward which the models converge at high mass is the ZAMS; by 3 Myr all stars above $\sim$ 2 $\rm M_\odot$ have reached it. At smaller masses, the maximum radius occurs at a mass that corresponds to stars that have just made the convective-radiative core transition at a given time. This value shifts to progressively smaller masses at later times.

An interesting feature of Figure~\ref{fig:varytdisk} is that the stellar rotation rates as a fraction of break up show a bimodal distribution: stars with $\rm M_{\star, f} \lesssim 1 \: M_\odot$  rotate at $\sim$10\% of their break up speed whereas stars with $\rm M_{\star, f} \gtrsim 6 \: M_\odot$ are rapid rotators. In between these plateaus (i.e., the ``transition region") the rotation rates as a fraction of break up increases with stellar mass. Furthermore, as time increases we find that the ratio of rotation speed to break up speed decreases on both plateaus, but that this decrease is more noticeable for the fast rotator plateau. This is because the stars located on the fast rotator plateau have already reached the ZAMS and are no longer contracting whereas those located on the slow rotator plateau are easy to spin down because of their low inertia, even though they are still contracting towards the ZAMS. In contrast,  we find that the rotation rates as a fraction of break up of the stars in the transition region increases with time. This suggests that the magnetic torques conveyed by the star-disk interaction are unable to counteract the increase in the stellar spin rate due to contraction for stars in the transition region. However, once these stars have reached the ZAMS magnetic torques do become important. The points located in the bottom right panel of Figure~\ref{fig:varytdisk} represents the minimum mass of stars rotating at $\gtrsim$20\% of their break up speed. We use this as an indicator of the transition between slow and fast rotators, which we discuss further in $\S$4.3. 

\begin{figure}[!t]
\centering
\includegraphics[angle = 90, trim=0cm 0cm 2cm 0cm, clip=true, width=0.52\textwidth]{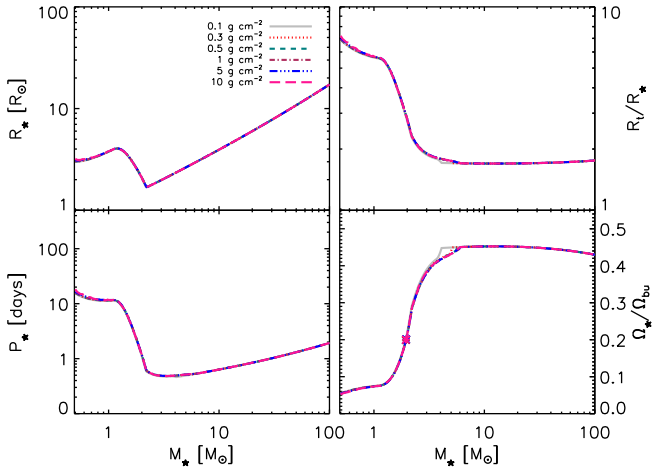}
\caption{Same as figure 4 but all quantities are shown at a time of 3 Myr, and we vary $\Sigma$ as indicated in the legend.}
\label{fig:varysigma}
\end{figure}

\subsection{Sensitivity to Model Parameters}
In the previous subsection we found that massive stars are much more difficult to spin down than low-mass stars. This causes low-mass stars to become slow rotators and massive stars to be rapid rotators, yielding a bimodal distribution in stellar rotation speeds as a fraction of the break up speed. To explore if this qualitative result is sensitive to our chosen model parameters, we vary certain parameters while holding the other parameters fixed. In the figures that follow we see that by varying certain parameters we do not lose this feature, but only alter it. 

\subsubsection{Varying $\Sigma$}
Figure~\ref{fig:varysigma} shows the final stellar radius, disk truncation radius, stellar period, and rotation rate as a fraction of break up as a function of final stellar mass for different values of the initial core surface density, $\Sigma$. The accretion rate during the main accretion phase increases for higher $\Sigma$, so varying this value affects the accretion history only during this phase. We find that this parameter has little to no effect on the final spin rate of the stars because the magnetic torques are unimportant during this accretion phase. The very minor differences that do appear arise because the value of $\Sigma$ affects the time at which a star of a given final mass reaches the swelling phase: the swelling phase of the star occurs earlier in time at lower $\Sigma$. For each value of $\Sigma$ used in our models there is a slight kink in between  $\rm M_{\star, f} \approx  3-6 \: M_\odot$ and the location of this kink decreases in mass for smaller values of $\Sigma$. Stars to the right of this kink experience the jump in radius, discussed in $\S$4.1, before the end of the main accretion phase whereas those to the left experience the swelling during the disk-clearing accretion phase. However, the net effect on the stellar rotation rate is obviously minor.

\begin{figure}[!t]
\centering
\includegraphics[angle = 90, trim=0cm 0cm 2cm 0cm, clip=true, width=0.52\textwidth]{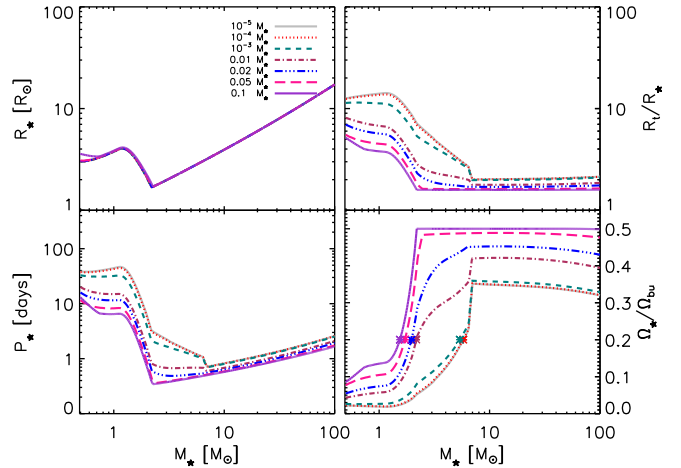}
\caption{Same as figure 4 but all quantities are shown at a time of 3 Myr, and we vary $M_{\rm D}$ as indicated in the legend.}
\label{fig:varyMD}
\end{figure}

\begin{figure}[!t]
\centering
\includegraphics[angle = 90, trim=0cm 0cm 2cm 0cm, clip=true, width=0.52\textwidth]{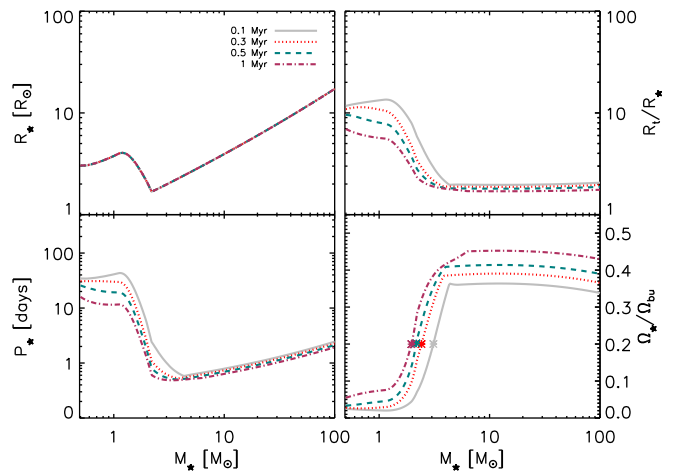}
\caption{Same as figure 4 but all quantities are shown at a time of 3 Myr, and we vary $t_{\rm a}$ as indicated in the legend.}
\label{fig:varytdecay}
\end{figure}

\subsubsection{Varying $M_{\rm D}$}
Figure~\ref{fig:varyMD} shows the final stellar radius, disk truncation radius, stellar period, and rotation rate as a fraction of break up as a function of final stellar mass for different values of the initial disk mass, $M_{\rm D}$, used for the disk clearing accretion phase. Increasing $M_{\rm D}$ increases the accretion rate during the disk clearing phase, thus increasing the accretion torque. A larger accretion rate also causes the disk to be truncated closer to the star, effectively reducing the net spin down magnetic torque. This is because the stellar magnetic field lines will connect to a greater portion of the disk within $R_{\rm co}$ yielding greater spin up magnetic torques on the star while the magnetic spin down torques remain unchanged. We find that altering $M_{\rm D}$ changes the location and shape of the transition between the slow and fast rotation plateaus, but the qualitative result that rotation rates are bimodal, with slow rotation at low mass and rapid rotation at high mass, remains unchanged. Also note that the models converge in the limit $M_{\rm D} \rightarrow 0$.

\subsubsection{Varying $t_{\rm a}$}
Figure~\ref{fig:varytdecay}  shows the final stellar radius, disk truncation radius, stellar period, and rotation rate as a fraction of break up as a function of final stellar mass for different values of the disk decay time scale, $t_{\rm a}$, used for equation (16) . Smaller values of $t_{\rm a}$, as compared to our fiducial value of 1 Myr, correspond to a higher initial accretion rate that declines more rapidly for the disk clearing accretion phase. This yields lower final spin rates at the end of 3 Myr. However, the overall shape of the distribution of final spin rates as a function of stellar mass does not change.

\subsubsection{Varying $B_{\rm \star}$}
\begin{figure}[!t]
\centering
\includegraphics[angle = 90, trim=0cm 0cm 2cm 0cm, clip=true, width=0.52\textwidth]{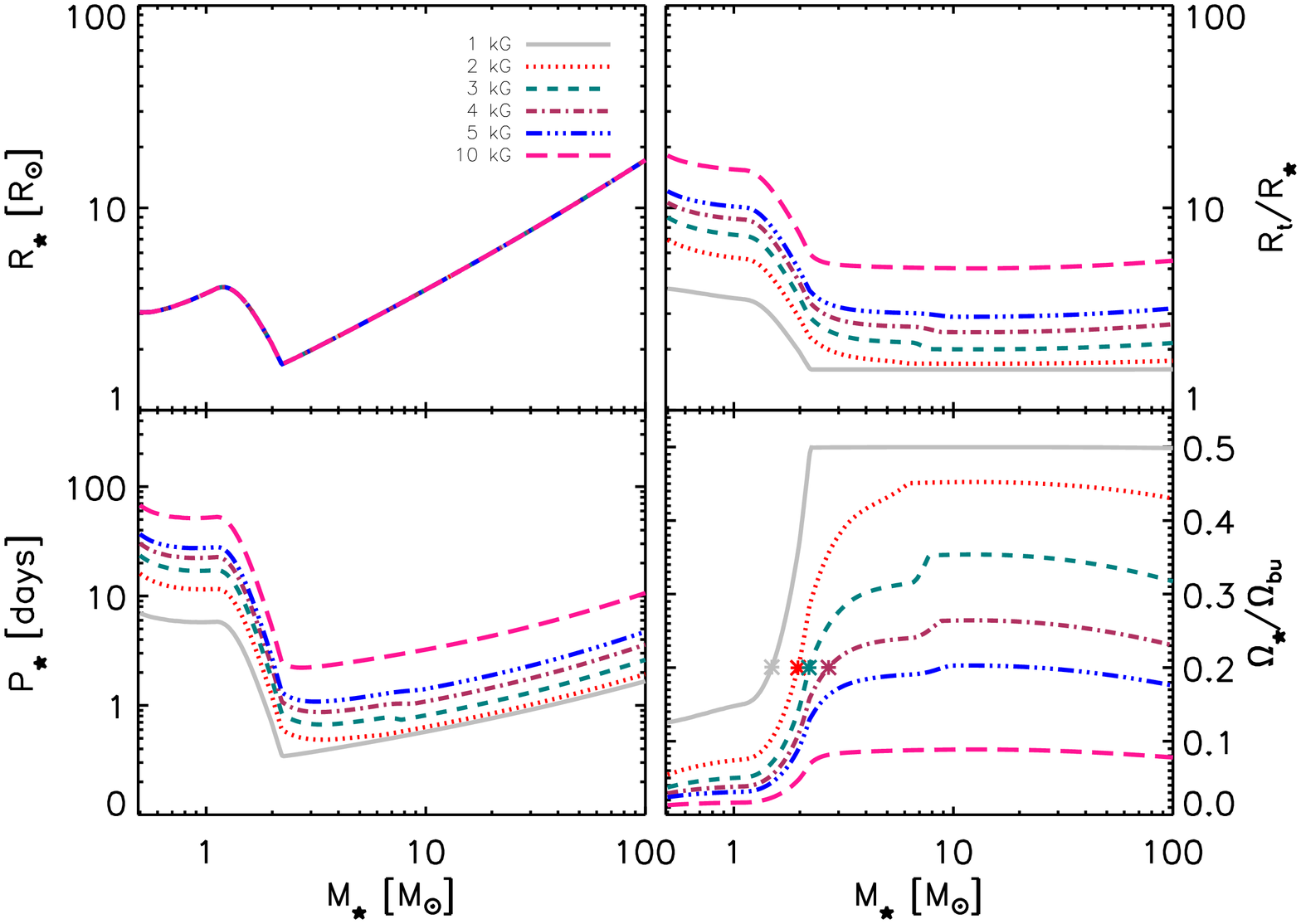}
\caption{Same as figure 4 but all quantities are shown at a time of 3 Myr, and we vary $B_{\rm \star}$ as indicated in the legend.}
\label{fig:varyB}
\end{figure}

Figure~\ref{fig:varyB}  shows the final stellar radius, disk truncation radius, stellar period, and rotation rate as a fraction of break up as a function of final stellar mass for different values of the stellar magnetic field strength. Clearly, a larger magnetic field strength provides a greater spin down torque on the star, yielding smaller final spin rates as a function of mass. As can be seen in this figure, stars above  $\rm M_{\star, f} \gtrsim 2 \: M_\odot$ require surface fields greater than 1 kG to experience any significant spin down torques and do not become slow rotators, $\Omega_{\rm \star} / \Omega_{\rm bu} \lesssim 0.1$, unless the field reaches $\sim$10 kG. Magnetic fields this large have only been detected in the chemically peculiar (e.g., helium strong) Ap/Bp stars \citep{bl79,owm10}. Generally, as the field strength increases the final spin rates decrease, but the qualitative division between slow and fast rotators remains. We also find that this same trend in rotation rates as a fraction of break up occurs as the field lines become weakly coupled  to the accretion disk,  while holding the magnetic field strength fixed, as discussed next.

\subsubsection{Varying $\beta$ and $\gamma_{\rm c}$}
\begin{figure}[!t]
\centering
\includegraphics[angle = 90, trim=0cm 0cm 2cm 0cm, clip=true, width=0.52\textwidth]{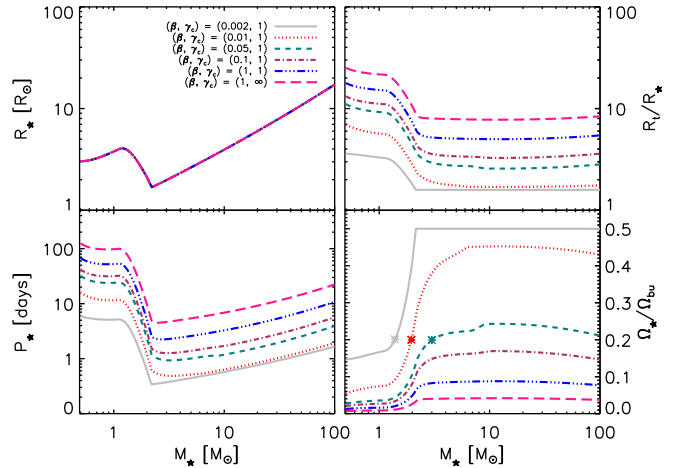}
\caption{Same as figure 4 but all quantities are shown at a time of 3 Myr, and we vary $\beta$ and $\gamma_c$ as indicated in the legend.}
\label{fig:varybeta}
\end{figure}

Figure~\ref{fig:varybeta}  shows the final stellar radius, disk truncation radius, stellar period, and rotation rate as a fraction of break up as a function of final stellar mass for different values of $\beta$ and $\gamma_{\rm c}$. These parameters describe the coupling and connection of the stellar magnetic field lines to the accretion disk (i.e., the location where the field lines open and disconnect from the disk). A larger $\beta$ for a given $\gamma_{\rm c}$ increases the extent of the connected disk region. This is because the coupling of the stellar field lines to the disk acts to resist the twisting of these lines due to the differential rotation between the star and disk. Thus, weaker field coupling will lead to a greater spin-down torque acting on the star leading to lower rotation rates as depicted in Figure~\ref{fig:varybeta}. Likewise, a greater $\gamma_{\rm c}$ for a given $\beta$ will allow the field lines to experience a greater twist before opening, also increasing the size of the connected disk region. For the case where $\gamma_{\rm c} \rightarrow \infty$ (i.e., field lines are allowed to twist to large values without opening), the field lines will connect to the whole disk outside $R_{\rm t}$. This will lead to a greater spin down torque. The case where $\beta$ = 1 and $\gamma_{\rm c}$ = $\infty$ reduces to the case described in section 2. Figure~\ref{fig:varybeta}  shows that as $\beta$ increases for $\gamma_{\rm c}=1$, all stars have lower rotation rates. However the two plateaus still remain.

\subsection{The Characteristic Mass for the Slow to Fast Rotator Transition}
In this work we have found a robust division between slow and fast rotators. Specifically, we find that low-mass stars (e.g., stars with $M_{\rm \star} \lesssim \rm{1\;M_\odot}$) are slow rotators, easily spun down via magnetic torques that arise from the star-disk interaction, and rotate at $\sim10\%$ of their break up speed, whereas massive stars (e.g., $M_{\rm \star} \gtrsim \rm{6\;M_\odot}$) are preferentially fast rotators. This is because massive stars are difficult to spin down due to their larger inertia and because their magnetic fields are weaker relative to their stellar binding energy as compared to low mass stars. Furthermore, this division is also dependent on the R-M relationship. The stars located on the fast rotator plateau have reached the ZAMS by the end of the main accretion phase or early on during the disk clearing phase; whereas, the stars located on the slow rotator plateau are shrinking towards the ZAMS for the entirety of the disk clearing phase. Likewise, the stars located in the transition region are contracting towards the ZAMS for a significant portion of the disk clearing phase but are contracting much faster than the low-mass slow rotators, leading to the sudden rise in rotation rates as a fraction of break up.

\begin{figure}[!t]
\centering
\includegraphics[width=0.50\textwidth]{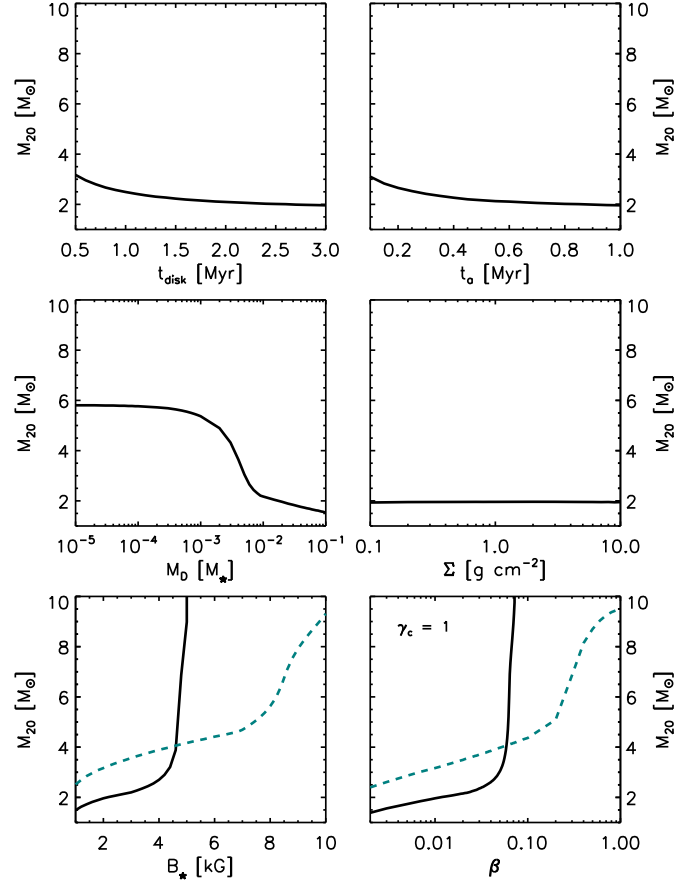}
\caption{This figure illustrates the sensitivity of the model parameters. The y-axes show the minimum stellar mass where $f\ge0.2$, denoted as $M_{\rm 20}$, for different parameters as indicated on the x-axes. Except for the top left plot, the black solid lines indicate that these values were taken for a disk lifetime of 3 Myrs. In the bottom panels, the teal dashed lines show the value of $M_{\rm 20}$ 0.5 Myrs after the beginning of the disk clearing phase for comparison.}
\label{fig:sum}
\end{figure}

To further illustrate the division between slow and fast rotators for each of our model parameters, in figure~\ref{fig:sum} we plot the minimum stellar mass at which the star ends accretion rotating at 20\% of its break up speed, which we call $M_{\rm 20}$. Each panel shows how  $M_{\rm 20}$ depends on the individual parameters in our model (while setting the other parameters to their fiducial values). The top panels show that  $M_{\rm 20}$ decreases by only a small amount as the disk lifetime (i.e., the amount of time the disk survives and supplies mass to the star during the disk clearing phase) or disk decay time scale increases. We also see that this characteristic mass, as a function of the initial core surface density, is relatively constant as indicated by the nearly horizontal line on the middle right panel of figure~\ref{fig:sum}. In contrast,  $M_{\rm 20}$ spans a larger mass range as we vary the initial disk mass used for the secondary accretion phase as shown in the middle left panel. This is because the accretion rate, and therefore the accretion torque, is proportional to the disk mass used in our model. We find that as $M_{\rm D} \rightarrow 0$ the values of $M_{\rm 20}$ become constant but we notice that $M_{\rm 20}$ decreases most as the disk mass increases from $\sim10^{-3}-10^{-2}\;M_{\rm \star}$. The division between slow and fast rotators slowly decreases in stellar mass for initial disk masses above $\sim 10^{-2}\;M_{\rm \star}$. Even though varying this parameter leads to larger variations in $M_{\rm 20}$ as compared to the top panels, it does not change the qualitative division between slow and fast rotators. 
 
The bottom panels of figure~\ref{fig:sum} show how the slow-fast rotator division is affected by the stellar magnetic field strength and the coupling of the stellar magnetic field lines to the disk, which are the parameters that are responsible for the removal of angular momentum from the star. The black solid lines in these panels show that the division between the slow and fast rotators (i.e., $M_{\rm 20}$) diverges for large magnetic field strengths ($B_{\rm \star} \gtrsim$ 4.5 kG) or weak field coupling ($\beta \gtrsim$ 0.05) for a disk clearing accretion phase that lasts for 3 Myr. This is because no stars will be rotating at or above 20\% of their break up speed at the end of 3 Myr for such high values of $B_{\rm \star}$ or $\beta$. For comparison, and also because we expect disks to have shorter lifetimes around massive stars, we also include the values of $M_{\rm 20}$ at 0.5 Myr after the disk clearing phase began (teal dotted lines). We find that $M_{\rm 20}$ is larger at shorter times because these stars are still contracting towards the ZAMS. At 0.5 Myr stars with masses greater than $\sim 5$ $\rm M_\odot$ have reached the ZAMS, as indicated by the kink and faster increase of $M_{\rm 20}$ in these plots for the 0.5 Myr case. 

\section{Discussion}
We have shown that massive stars are fast rotators at birth and that their initial rotation rates are unlikely to be regulated by the star-disk magnetic interaction. We have found that magnetic torques can only effectively spin down massive stars that have low accretion rates, long disk lifetimes, weak magnetic coupling with the disk, and/or surface magnetic fields that are significantly larger than what current observational estimates suggest. We thus conclude that their initial rotation rates are likely regulated by gravitational torques. Since massive stars arrive on the main sequence as fast rotators, their variation in rotation rates as a fraction of their break-up rate is likely a result of evolutionary spin down, due to stellar expansion and/or angular momentum loss via stellar winds while on the main sequence. 

\subsection{Observational Implications}
A topic of current debate is whether the distribution of the projected rotational velocities of massive stars depend on birth environment or if this property is only affected by evolutionary spin down \citep{swd05, dsl06, hg06,hg08, ws07, ws08, hgm10}. \citet{swd05} observed the rotational velocities of B stars located in high stellar density clusters and compared them to field stars of similar age ($\sim$12-15 Myr). They found that, on average, the cluster stars had larger rotational velocities than the field stars in their sample and that only the most evolved cluster stars had similar rotational velocities as their field star counterparts. Likewise, \citet{ws07,ws08} observed that massive stars (e.g., $M_{\rm \star}\gtrsim\;6\; \rm M_\odot$) found in clusters characterized by a high stellar density are faster rotators than their similar mass counterparts located in lower density clusters. These studies concluded that the initial spin rates of these stars depend on the initial star-forming environment since these stellar ensembles, which have survived as bound clusters, likely form in molecular clouds characterized by high surface densities.  Furthermore, \citet{ws07} compared the distribution of the rotational velocities of B stars in both young and older high density and low density environments and did not detect a significant evolutionary change. 

In agreement, \citet{hgm10} compared the rotation rates of cluster and field B stars and found that, on average, cluster stars tend to rotate faster than field stars. However, by grouping the stars by surface gravity, an age proxy, they found  there is little difference between the average rotational velocities for the field and cluster stars as a function of age, and that they exhibit a similar spin-down with advanced evolution. They also found that field stars are in general more evolved than cluster stars. These results suggest that the observed trend in the rotational velocities of B stars are due to evolutionary spin down rather than to the initial conditions of the environment in which they formed. They argue that the discrepancy between the average rotation rate of the field stars and cluster stars in their sample is that the field stars have undergone evolutionary spin down since the field star sample contained more evolved stars.

For a fixed surface magnetic field strength, we find here that the initial rotation rates of massive stars, due to disk locking, have no dependence on the environmental density. As described in $\S$3.2.1, the accretion rate during the main accretion phase does depend on the star forming environment, with larger surface density yielding a greater time-averaged accretion rate. \citet{ws07} proposed that the higher rotation rates they report for stars in dense clusters are the result of disk-locking plus a systematically higher accretion rate in dense clusters. However, we find that magnetic torques are insignificant during the main accretion phase regardless of environment density because of the high accretion rates. These torques only become important during the disk clearing phase, and there is no obvious reason that the properties or behavior of the disk during this phase should depend on the environment. However, this does not rule out other factors that may depend on the environment. In this work we assumed that all stars had the same surface magnetic field strength. If the strength of the magnetic fields present during the star formation process depends on environment, either because the star-forming cloud has a different magnetic mass to flux ratio and/or because the ambipolar diffusion process depends on density, then this could provide a viable explanation for the difference in rotational velocities of young stars in environments of varying density. Another possibility for the difference in rotational velocities of stars born in different environments may be related to the lifetimes of disks in such environments. We have found that the rotation rates of these stars depend crucially on the lifetime of the accretion disk. Thus, if disks have shorter lifetimes in higher stellar density environments, possibly due to tidal dissipation from interactions with neighbors or rapid photoevaporation due to radiation from nearby massive stars, then the initial rotation rates of these stars will only increase as they contract towards the ZAMS \citep{ws07}.
 
 \subsection{Future Work and Caveats}
 
In this work, we have omitted two potentially important effects: that magnetic fields might be stronger early in stars' lives, and that stars can be spun down by winds on the main sequence. As mentioned in $\S$1, magnetic fields in massive stars are likely to be the decaying remnants of magnetic flux swept up during the star formation process. Therefore, it is plausible that accreting massive stars have stronger magnetic fields than those we observe as main sequence O and B stars. If this is the case, then massive stars will likely be spun-down via magnetic torques. If the decay process is the same for all stars then we expect that the strongest magnetic fields should be observed in the slowest rotators. However, we also discovered that the spin rates of these stars depend heavily on how well the stellar magnetic field lines couple to the accretion disk. As described in $\S$3.3.1 the true value of $\beta$ is highly uncertain because it depends on the microphysics of the accretion disk. Since observations of disks around massive stars are rare we are unable to provide a confident estimate for $\beta$. However, by exploring a range of values for $\beta$ we have determined that if the field lines are weakly coupled to the disk then magnetic torques can sufficiently spin down massive stars. Also, measuring the rotation rates of young, massive stars can provide a better estimate for $\beta$. If the slowest rotators prove to have weak magnetic fields, then it may be likely that the field lines were weakly coupled to the disk, resulting in a larger $\beta$, thus producing these slower rotators.

Stars on the main sequence also shed mass and angular momentum via stellar winds, which we have neglected in this work. In the presence of a stellar magnetic field, these winds will couple with the field lines causing the star to lose a significant amount of angular momentum as it evolves \citep{wd67}. Since the mass loss rates of stars increases with stellar mass \citep{nie90} more massive stars will lose angular momentum at a greater rate. If the spin rates of massive stars are regulated by gravitational torques rather than magnetic torques produced by the star-disk magnetic interaction then we expect that all massive stars should be rotating at $\sim$50\% of their break up speed once they are deposited on the ZAMS, assuming that their disks survive long enough. Spin down will occur as they evolve and shed angular momentum via stellar winds. This is consistent with the results of \citet{hgm10} who found that young stars with masses greater than $\sim$ 2 $M_{\rm \odot}$ are preferentially fast rotators and that the average rotation speed as a fraction of the break up speed, for each mass bin, decreases for increasing stellar mass. 

\acknowledgements{ALR acknowledges support from the NSF GSFP. MRK is supported by the Alfred P. Sloan Foundation, the NSF through grant CAREER-0955300, and NASA through grant NNX09AK31G and a Chandra Telescope grant. ERR acknowledges support from the David and Lucille Packard Foundation and NSF grant: AST-0847563. ALR thanks Nathan Goldbaum, James Guillochon, Nic Ross, and Rachel Strickler for useful discussions. We thank the editor and the referee for taking the time to carefully read our manuscript and provide useful comments.}

\bibliographystyle{apj}
\bibliography{anna2011_bib}

\begin{thebibliography}{65}
\expandafter\ifx\csname natexlab\endcsname\relax\def\natexlab#1{#1}\fi

\bibitem[{{Alecian} {et~al.}(2008){Alecian}, {Wade}, {Catala}, {Bagnulo},
  {Boehm}, {Bohlender}, {Bouret}, {Donati}, {Folsom}, {Grunhut}, \&
  {Landstreet}}]{awc08}
{Alecian}, E., {et~al.} 2008, \aap, 481, L99

\bibitem[{{Armitage} \& {Clarke}(1996)}]{ac96}
{Armitage}, P.~J., \& {Clarke}, C.~J. 1996, \mnras, 280, 458

\bibitem[{{Beuther} {et~al.}(2002){Beuther}, {Schilke}, {Menten}, {Motte},
  {Sridharan}, \& {Wyrowski}}]{bsmm02}
{Beuther}, H., {Schilke}, P., {Menten}, K.~M., {Motte}, F., {Sridharan}, T.~K.,
  \& {Wyrowski}, F. 2002, \apj, 566, 945

\bibitem[{{Bjorkman} \& {Cassinelli}(1993)}]{bc93}
{Bjorkman}, J.~E., \& {Cassinelli}, J.~P. 1993, \apj, 409, 429

\bibitem[{{Bodenheimer}(1995)}]{bod95}
{Bodenheimer}, P. 1995, \araa, 33, 199

\bibitem[{{Borra} \& {Landstreet}(1979)}]{bl79}
{Borra}, E.~F., \& {Landstreet}, J.~D. 1979, \apj, 228, 809

\bibitem[{{Bouvier}(2007)}]{bou07}
{Bouvier}, J. 2007, in IAU Symposium, Vol. 243, IAU Symposium, ed. {J.~Bouvier
  \& I.~Appenzeller}, 231--240

\bibitem[{{Caselli} \& {Myers}(1995)}]{cm95}
{Caselli}, P., \& {Myers}, P.~C. 1995, \apj, 446, 665

\bibitem[{{Cesaroni} {et~al.}(2007){Cesaroni}, {Galli}, {Lodato}, {Walmsley},
  \& {Zhang}}]{cglrev07}
{Cesaroni}, R., {Galli}, D., {Lodato}, G., {Walmsley}, C.~M., \& {Zhang}, Q.
  2007, Protostars and Planets V, 197

\bibitem[{{Cesaroni} {et~al.}(2006){Cesaroni}, {Galli}, {Lodato}, {Walmsley},
  \& {Zhang}}]{cesnat06}
{Cesaroni}, R., {Galli}, D., {Lodato}, G., {Walmsley}, M., \& {Zhang}, Q. 2006,
  \nat, 444, 703

\bibitem[{{Chini} {et~al.}(2006){Chini}, {Hoffmeister}, {Nielbock}, {Scheyda},
  {Steinacker}, {Siebenmorgen}, \& {N{\"u}rnberger}}]{chi06}
{Chini}, R., {Hoffmeister}, V.~H., {Nielbock}, M., {Scheyda}, C.~M.,
  {Steinacker}, J., {Siebenmorgen}, R., \& {N{\"u}rnberger}, D. 2006, \apjl,
  645, L61

\bibitem[{{Chini} {et~al.}(2011){Chini}, {Hoffmeister}, \&
  {N{\"u}rnberger}}]{chn11}
{Chini}, R., {Hoffmeister}, V.~H., \& {N{\"u}rnberger}, D. 2011, Bulletin de la
  Societe Royale des Sciences de Liege, 80, 217

\bibitem[{{Collier Cameron} \& {Campbell}(1993)}]{cc93}
{Collier Cameron}, A., \& {Campbell}, C.~G. 1993, \aap, 274, 309

\bibitem[{{Crutcher}(1999)}]{cru99}
{Crutcher}, R.~M. 1999, \apj, 520, 706

\bibitem[{{Cunningham} {et~al.}(2011){Cunningham}, {Klein}, {Krumholz}, \&
  {McKee}}]{ckk11}
{Cunningham}, A.~J., {Klein}, R.~I., {Krumholz}, M.~R., \& {McKee}, C.~F. 2011,
  \apj, 740, 107

\bibitem[{{Davies} {et~al.}(2011){Davies}, {Hoare}, {Lumsden}, {Hosokawa},
  {Oudmaijer}, {Urquhart}, {Mottram}, \& {Stead}}]{dav11}
{Davies}, B., {Hoare}, M.~G., {Lumsden}, S.~L., {Hosokawa}, T., {Oudmaijer},
  R.~D., {Urquhart}, J.~S., {Mottram}, J.~C., \& {Stead}, J. 2011, \mnras, 1015

\bibitem[{{Donati} {et~al.}(2006){Donati}, {Howarth}, {Bouret}, {Petit},
  {Catala}, \& {Landstreet}}]{don02}
{Donati}, J.-F., {Howarth}, I.~D., {Bouret}, J.-C., {Petit}, P., {Catala}, C.,
  \& {Landstreet}, J. 2006, \mnras, 365, L6

\bibitem[{{Dufton} {et~al.}(2006){Dufton}, {Smartt}, {Lee}, {Ryans}, {Hunter},
  {Evans}, {Herrero}, {Trundle}, {Lennon}, {Irwin}, \& {Kaufer}}]{dsl06}
{Dufton}, P.~L., {et~al.} 2006, \aap, 457, 265

\bibitem[{{Ghosh} \& {Lamb}(1978)}]{gl78}
{Ghosh}, P., \& {Lamb}, F.~K. 1978, \apjl, 223, L83

\bibitem[{{Goodman} {et~al.}(1993){Goodman}, {Benson}, {Fuller}, \&
  {Myers}}]{good93}
{Goodman}, A.~A., {Benson}, P.~J., {Fuller}, G.~A., \& {Myers}, P.~C. 1993,
  \apj, 406, 528

\bibitem[{{Grunhut} {et~al.}(2009){Grunhut}, {Wade}, {Marcolino}, {Petit},
  {Henrichs}, {Cohen}, {Alecian}, {Bohlender}, {Bouret}, {Kochukhov}, {Neiner},
  {St-Louis}, \& {Townsend}}]{gru09}
{Grunhut}, J.~H., {et~al.} 2009, \mnras, 400, L94

\bibitem[{{Hartmann} {et~al.}(2006){Hartmann}, {D'Alessio}, {Calvet}, \&
  {Muzerolle}}]{hdcm06}
{Hartmann}, L., {D'Alessio}, P., {Calvet}, N., \& {Muzerolle}, J. 2006, \apj,
  648, 484

\bibitem[{{Hartmann} \& {Stauffer}(1989)}]{har89}
{Hartmann}, L., \& {Stauffer}, J.~R. 1989, \aj, 97, 873

\bibitem[{{Herbst} {et~al.}(2007){Herbst}, {Eisl{\"o}ffel}, {Mundt}, \&
  {Scholz}}]{hem07}
{Herbst}, W., {Eisl{\"o}ffel}, J., {Mundt}, R., \& {Scholz}, A. 2007,
  Protostars and Planets V, 297

\bibitem[{{Hosokawa} {et~al.}(2011){Hosokawa}, {Offner}, \& {Krumholz}}]{hok11}
{Hosokawa}, T., {Offner}, S.~S.~R., \& {Krumholz}, M.~R. 2011, \apj, 738, 140

\bibitem[{{Hosokawa} \& {Omukai}(2009)}]{ho09}
{Hosokawa}, T., \& {Omukai}, K. 2009, \apj, 691, 823

\bibitem[{{Huang} \& {Gies}(2006)}]{hg06}
{Huang}, W., \& {Gies}, D.~R. 2006, \apj, 648, 580

\bibitem[{{Huang} \& {Gies}(2008)}]{hg08}
---. 2008, \apj, 683, 1045

\bibitem[{{Huang} {et~al.}(2010){Huang}, {Gies}, \& {McSwain}}]{hgm10}
{Huang}, W., {Gies}, D.~R., \& {McSwain}, M.~V. 2010, \apj, 722, 605

\bibitem[{{Hubrig} {et~al.}(2008){Hubrig}, {Sch{\"o}ller}, {Schnerr},
  {Gonz{\'a}lez}, {Ignace}, \& {Henrichs}}]{hub08}
{Hubrig}, S., {Sch{\"o}ller}, M., {Schnerr}, R.~S., {Gonz{\'a}lez}, J.~F.,
  {Ignace}, R., \& {Henrichs}, H.~F. 2008, \aap, 490, 793

\bibitem[{{Johns-Krull}(2007)}]{jk07}
{Johns-Krull}, C.~M. 2007, \apj, 664, 975

\bibitem[{{Koenigl}(1991)}]{koe91}
{Koenigl}, A. 1991, \apjl, 370, L39

\bibitem[{{Krumholz} {et~al.}(2007){Krumholz}, {Klein}, \& {McKee}}]{krum07}
{Krumholz}, M.~R., {Klein}, R.~I., \& {McKee}, C.~F. 2007, \apj, 656, 959

\bibitem[{{Krumholz} {et~al.}(2009){Krumholz}, {Klein}, {McKee}, {Offner}, \&
  {Cunningham}}]{krum09}
{Krumholz}, M.~R., {Klein}, R.~I., {McKee}, C.~F., {Offner}, S.~S.~R., \&
  {Cunningham}, A.~J. 2009, Science, 323, 754

\bibitem[{{Larson}(2010)}]{lar10}
{Larson}, R.~B. 2010, Reports on Progress in Physics, 73, 014901

\bibitem[{{Lin} {et~al.}(2011){Lin}, {Krumholz}, \& {Kratter}}]{lkk11}
{Lin}, M.-K., {Krumholz}, M.~R., \& {Kratter}, K.~M. 2011, \mnras, 1047

\bibitem[{{Lovelace} {et~al.}(1995){Lovelace}, {Romanova}, \&
  {Bisnovatyi-Kogan}}]{love95}
{Lovelace}, R.~V.~E., {Romanova}, M.~M., \& {Bisnovatyi-Kogan}, G.~S. 1995,
  \mnras, 275, 244

\bibitem[{{Maeder} \& {Meynet}(2010)}]{mm10}
{Maeder}, A., \& {Meynet}, G. 2010, \nar, 54, 32

\bibitem[{{Martins} {et~al.}(2010){Martins}, {Donati}, {Marcolino}, {Bouret},
  {Wade}, {Escolano}, \& {Howarth}}]{mar10}
{Martins}, F., {Donati}, J.-F., {Marcolino}, W.~L.~F., {Bouret}, J.-C., {Wade},
  G.~A., {Escolano}, C., \& {Howarth}, I.~D. 2010, \mnras, 407, 1423

\bibitem[{{Matt} \& {Pudritz}(2005)}]{mp05a}
{Matt}, S., \& {Pudritz}, R.~E. 2005, \mnras, 356, 167

\bibitem[{{Matt} {et~al.}(2010){Matt}, {Pinz{\'o}n}, {de la Reza}, \&
  {Greene}}]{matt10}
{Matt}, S.~P., {Pinz{\'o}n}, G., {de la Reza}, R., \& {Greene}, T.~P. 2010,
  \apj, 714, 989

\bibitem[{{Matzner} \& {McKee}(2000)}]{mm00}
{Matzner}, C.~D., \& {McKee}, C.~F. 2000, \apj, 545, 364

\bibitem[{{McKee} \& {Tan}(2003)}]{mt03}
{McKee}, C.~F., \& {Tan}, J.~C. 2003, \apj, 585, 850

\bibitem[{{Moss}(2001)}]{moss01}
{Moss}, D. 2001, in Astronomical Society of the Pacific Conference Series, Vol.
  248, Magnetic Fields Across the Hertzsprung-Russell Diagram, ed. {G.~Mathys,
  S.~K.~Solanki, \& D.~T.~Wickramasinghe}, 305--+

\bibitem[{{Myers} \& {Fuller}(1992)}]{mf92}
{Myers}, P.~C., \& {Fuller}, G.~A. 1992, \apj, 396, 631

\bibitem[{{Nakano} {et~al.}(2000){Nakano}, {Hasegawa}, {Morino}, \&
  {Yamashita}}]{nhmy00}
{Nakano}, T., {Hasegawa}, T., {Morino}, J.-I., \& {Yamashita}, T. 2000, \apj,
  534, 976

\bibitem[{{Nieuwenhuijzen} \& {de Jager}(1990)}]{nie90}
{Nieuwenhuijzen}, H., \& {de Jager}, C. 1990, \aap, 231, 134

\bibitem[{{Offner} {et~al.}(2009){Offner}, {Klein}, {McKee}, \&
  {Krumholz}}]{off09}
{Offner}, S.~S.~R., {Klein}, R.~I., {McKee}, C.~F., \& {Krumholz}, M.~R. 2009,
  \apj, 703, 131

\bibitem[{{Oksala} {et~al.}(2010){Oksala}, {Wade}, {Marcolino}, {Grunhut},
  {Bohlender}, {Manset}, {Townsend}, \& {Mimes Collaboration}}]{owm10}
{Oksala}, M.~E., {Wade}, G.~A., {Marcolino}, W.~L.~F., {Grunhut}, J.,
  {Bohlender}, D., {Manset}, N., {Townsend}, R.~H.~D., \& {Mimes
  Collaboration}. 2010, \mnras, 405, L51

\bibitem[{{Parravano} {et~al.}(2003){Parravano}, {Hollenbach}, \&
  {McKee}}]{phm03}
{Parravano}, A., {Hollenbach}, D.~J., \& {McKee}, C.~F. 2003, \apj, 584, 797

\bibitem[{{Press} {et~al.}(2007){Press}, {Teukolsky}, {Vetterling}, \&
  {Flannery}}]{nr07}
{Press}, W.~H., {Teukolsky}, S.~A., {Vetterling}, W.~T., \& {Flannery}, B.~P.
  2007, Numerical Recipes: The Art of Scientific Computing (Cambridge
  University Press)

\bibitem[{{Shu}(1977)}]{shu77}
{Shu}, F.~H. 1977, \apj, 214, 488

\bibitem[{{Strom} {et~al.}(2005){Strom}, {Wolff}, \& {Dror}}]{swd05}
{Strom}, S.~E., {Wolff}, S.~C., \& {Dror}, D.~H.~A. 2005, \aj, 129, 809

\bibitem[{{Uzdensky} {et~al.}(2002){Uzdensky}, {K{\"o}nigl}, \&
  {Litwin}}]{ukl02}
{Uzdensky}, D.~A., {K{\"o}nigl}, A., \& {Litwin}, C. 2002, \apj, 565, 1191

\bibitem[{{van der Tak} {et~al.}(2000){van der Tak}, {van Dishoeck}, {Evans},
  \& {Blake}}]{vdt00}
{van der Tak}, F.~F.~S., {van Dishoeck}, E.~F., {Evans}, II, N.~J., \& {Blake},
  G.~A. 2000, \apj, 537, 283

\bibitem[{{Vlemmings} {et~al.}(2010){Vlemmings}, {Surcis}, {Torstensson}, \&
  {van Langevelde}}]{vsc10}
{Vlemmings}, W.~H.~T., {Surcis}, G., {Torstensson}, K.~J.~E., \& {van
  Langevelde}, H.~J. 2010, \mnras, 404, 134

\bibitem[{{Wade} {et~al.}(2006){Wade}, {Fullerton}, {Donati}, {Landstreet},
  {Petit}, \& {Strasser}}]{wade06}
{Wade}, G.~A., {Fullerton}, A.~W., {Donati}, J.-F., {Landstreet}, J.~D.,
  {Petit}, P., \& {Strasser}, S. 2006, \aap, 451, 195

\bibitem[{{Walder} {et~al.}(2011){Walder}, {Folini}, \& {Meynet}}]{wfm11}
{Walder}, R., {Folini}, D., \& {Meynet}, G. 2011, \ssr, 57

\bibitem[{{Weber} \& {Davis}(1967)}]{wd67}
{Weber}, E.~J., \& {Davis}, Jr., L. 1967, \apj, 148, 217

\bibitem[{{Wolff} {et~al.}(2008){Wolff}, {Strom}, {Cunha}, {Daflon}, {Olsen},
  \& {Dror}}]{ws08}
{Wolff}, S.~C., {Strom}, S.~E., {Cunha}, K., {Daflon}, S., {Olsen}, K., \&
  {Dror}, D. 2008, \aj, 136, 1049

\bibitem[{{Wolff} {et~al.}(2006){Wolff}, {Strom}, {Dror}, {Lanz}, \&
  {Venn}}]{ws06}
{Wolff}, S.~C., {Strom}, S.~E., {Dror}, D., {Lanz}, L., \& {Venn}, K. 2006,
  \aj, 132, 749

\bibitem[{{Wolff} {et~al.}(2007){Wolff}, {Strom}, {Dror}, \& {Venn}}]{ws07}
{Wolff}, S.~C., {Strom}, S.~E., {Dror}, D., \& {Venn}, K. 2007, \aj, 133, 1092

\bibitem[{{Yi}(1994)}]{yi94}
{Yi}, I. 1994, \apj, 428, 760

\bibitem[{{Yi}(1995)}]{yi95}
---. 1995, \apj, 442, 768

\bibitem[{{Zapata} {et~al.}(2008){Zapata}, {Palau}, {Ho}, {Schilke}, {Garrod},
  {Rodr{\'{\i}}guez}, \& {Menten}}]{zph08}
{Zapata}, L.~A., {Palau}, A., {Ho}, P.~T.~P., {Schilke}, P., {Garrod}, R.~T.,
  {Rodr{\'{\i}}guez}, L.~F., \& {Menten}, K. 2008, \aap, 479, L25

\end{thebibliography}

\end{document}